\documentclass{ws-procs961x669} 

\usepackage{amsmath,amsfonts,amssymb,latexsym,graphicx}

\def\be{\begin{equation}}
\def\ee{\end{equation}}
\def\bq{\begin{eqnarray}}
\def\eq{\end{eqnarray}}
\def\beq{\begin{eqnarray}}
\def\eeq{\end{eqnarray}}

\def\a{\alpha}

\def\t{\theta}

\begin{document}
\title{Primordial synchronization of Mixmaster spatial points}

\author{Spiros Cotsakis}

\address{Institute of Gravitation and Cosmology, RUDN University\\
ul. Miklukho-Maklaya 6, Moscow 117198, Russia\\ and\\
Research Laboratory of Geometry,  Dynamical Systems  and Cosmology\\
University of the Aegean, Karlovassi 83200, Samos, Greece\\
E-mail: skot@aegean.gr}

\begin{abstract}
We review recent work on the possibility of primordial synchronization of different Mixmaster regions in generic inhomogeneous spacetime. It is shown that inhomogeneous domains undergoing chaotic oscillations may synchronize themselves exponentially fast and proceed in perfect symphony asymptotically in the past direction. Implications of this result for the structure and evolution of an early phase of the universe are briefly discussed.
\end{abstract}

\keywords{Nature of the cosmological singularity; structure of primordial states; chaotic synchronization}

\bodymatter
\section{Introduction}
Recently, we introduced a dynamical mechanism via chaotic synchronization of different spatial point of Mixmaster type as a new factor governing the global structure and evolution of primordial cosmology \cite{c1}. We have shown that a pair of  spatial Mixmaster points in generic inhomogeneous spacetime  coupled to each other  and one of them (`the transmitter')  communicating to the other (`the receiver') only part of the information necessary to determine its state, evolves in such a way that the receiver becomes able to completely reconstruct the remaining information and  synchronize its chaotic evolution with that of the transmitter point.

This mechanism radically changes the structure of the early universe and provides a novel setting for the consideration of a number of fundamental issues in cosmology, such as the homogeneity (`horizon') problem, and the unexplained behaviour of the entropy in the early universe. In this contribution, we shall briefly recall the main features of the proposed cosmological sync mechanism \cite{c1}.

\section{Evolution of inhomogeneous Mixmaster points}
Our results concern  the so-called $G_0$ cosmologies that admit no isometries and may be described in vacuum by the dimensionless state vector field $X$ along the orbit of a fixed spatial point $A$. This is given by\cite{uvwe},
\be
X_A (\tau)=(\Sigma_+,\Sigma_-,N_1,N_2,N_3),\quad\Sigma_\pm=\frac{\sigma_\pm}{H},\,\,N_\a=\frac{n_\a}{H},
\ee
where $H$ denotes the Hubble scalar function equal to $1/3\Theta$ and is taken with respect to the fundamental 4-velocity timelike vector field $u^a, a=0,1,2,3$.  The expansion scalar is defined by $\Theta=\nabla_a u^a$. Derivatives are taken with respect to the dimensionless $\tau$-time related to the proper (clock) time by $dt/d\tau=1/H$,  the variables $\Sigma_\pm$ describe the anisotropy in the Hubble flow,  while the $N_\a$'s  are related to the spatial curvature and the Bianchi type of the isometry group associated with the evolution of the spatial point.

We assume that different spatial points evolve in the past direction as separate homogeneous Mixmaster universes coupled to each other in inhomogeneous spacetime. For any pair of spatial points $A,B$, we have a pair of dynamical equations, namely, for the point $A$ we have equations for the $(N,\Sigma)$ variables,  and for the point $B$  similar equations for the different $(M,\Pi)$ variables as follows\cite{em,wh,we}:

System $A$:
\bq
N'_1&=& (q-4\Sigma_+)N_1,\label{n1}\\
N'_2 &=& (q+2\Sigma_+ + 2\sqrt{3}\Sigma_-)N_2, \label{n2}\\
N'_3 &=& (q+2\Sigma_+ - 2\sqrt{3}\Sigma_-)N_3,\label{n3} \\
\Sigma'_+ &=& -(2 - q)\Sigma_+ -3S_+, \label{+}\\
\Sigma'_-&=& -(2 - q)\Sigma_- - 3S_-,\label{-}
\eq
with the constraint,
\be\label{constraint}
\Sigma_+^2+\Sigma_-^2 +\frac{3}{2}\left(N_1^2+N_2^2+N_3^2 -2(N_1N_2+N_2N_3+N_3N_1)\right)=1,
\ee
where,
\bq
q&=&2(\Sigma_+^2+\Sigma_-^2),\\
S_+&=&\frac{1}{2}\left((N_2-N_3)^2-N_1(2N_1-N_2-N_3)\right),\label{splus}\\
S_-&=&\frac{\sqrt{3}}{2}(N_3-N_2)(N_1-N_2-N_3)\label{sminus},
\eq
and a prime denoting differentiation with respect to the $\tau$-time.

System $B$: We have  the unknowns  $y=(M,\Pi)$,   the variables $M=(M_1,M_2,M_3)$ satisfy similar equations to the Eqns. (\ref{n1})-(\ref{n3}), and the shear variables $\Pi=(\Pi_+,\Pi_-)$ satisfy the system,
\bq
\Pi_{+}'&=&-(2-p)\Pi_+-3Q_+,\label{+pi}\\
\Pi_{-}'&=&-(2-p)\Pi_--3Q_-,\label{-pi}
\eq
with $p=2(\Pi_+^2+\Pi_-^2)$,  the $Q$'s are like the variables $S$'s in the Eqns. (\ref{splus}), (\ref{sminus}) but with the $M$'s in the corresponding places of $N$'s. The constraint is identical to (\ref{constraint}) but with the $(M,\Pi)$'s in the places of the $(N,\Sigma)$'s.

\section{The problem of cosmological sync}
We shall be interested in the possibility that the spatial points $A,B$, with $A$ in the causal past of $B$, can influence each other by sending signals from $A$ to $B$ and sync with each other in the past direction, while chaotically oscillating in inhomogeneous Mixmaster spacetime.

We imagine that the $A$-system  in the form $x'=f(x)$ for the variables $x=(N,\Sigma)$, breaks into two `subsystems',  with the $N$ variables satisfying the first three equations (`sys-1') and the $\Sigma$'s the last two (`sys-2'), namely,
\be
N'=g(N,\Sigma),\quad\Sigma'=h(N,\Sigma),
\ee
where $N=(N_1,N_2,N_3)$, $\Sigma=(\Sigma_+,\Sigma_-)$. We then set the $B$-system (receiver) variables $M$ simply equal to the corresponding $A$-transmitter variables $N$ (that is to those  in sys-1).  Then  the $B$-system  equations break into the pair consisting of the sys-1 equations, and those   identical to sys-2 but for the $\Pi$ variables, the `sys-3' system:
\bq
M_i&=&N_i,\quad i=1,2,3\label{13}\\
\Pi_{+}'&=&-(2-p)\Pi_+-3S_+,\label{+pi1}\\
\Pi_{-}'&=&-(2-p)\Pi_--3S_-.\label{-pi1}
\eq
Following  \cite{pc,pe97}, we  introduce the synchronization (error) function $\Omega=(\Omega_+,\Omega_-)$, which in the present case we take it to be: $\Omega=\Sigma-\Pi$, that is,
\be\label{omega}
\Omega_+=\Sigma_+-\Pi_+,\quad \Omega_-=\Sigma_--\Pi_-.
\ee
Then we say that there is complete synchronization of the two Mixmaster oscillating spatial points (that is of systems $A$ and $B$) provided we can show that,
\be\label{synch}
\Omega\rightarrow (0,0),\quad \textrm{as}\quad \tau\rightarrow-\infty.
\ee
If this happens, we say that system $A$ is synchronized with system $B$. 
Otherwise, the two oscillating spatial points corresponding to systems $A$ and $B$ will evolve autonomously, out of sync. 

In general,  we expect the appearance of extra coupling terms in the dynamical equations which describe the evolution of the sync function $\Omega$ during the past evolution the Mixmaster oscillating spatial points, the latter evolution given by Eqns. (\ref{+}), (\ref{-}), and (\ref{+pi}), (\ref{-pi}). The question then arises as to whether or not a system of spatial points chaotically oscillating in the past direction may sync and evolve in unison thereafter.

This is the problem of cosmological sync for the case of spatial points satisfying Mixmaster-type evolution equations. As shown in \cite{c1}, there are indeed different types of sync possible in this problem depending on the form of the coupling between different spatial points.

\section{Generic sync} 
Let us suppose that there is a linear in the sync function $\Omega$ coupling between the two spatial points. In this case, the  evolution equations become,
\bq
\Sigma' &=& -(2 - q)\Sigma -3S+\a(\Pi-\Sigma)\label{s}\\
\Pi' &=& -(2 - p)\Pi -3Q+\a(\Sigma-\Pi),\label{p}
\eq
where $Q$ is the analogous expression of $S$ for the $\Pi$ system, and $\a$ is some coupling between the two oscillators.

Then if the orbit $\Sigma$ has Lyapunov exponent $\lambda$, we find that the associated $\Omega$-evolution equation leads to the result\cite{c1},
\be \label{aa}
|\Omega|\leq Ce^{(\lambda-2\a)\tau},
\ee
with $C$ an integration constant.

Therefore if we set $\a_c=\lambda/2,$ for the critical coupling strength, we conclude that when
$
\a>\a_c,
$
we have complete synchronization of the Mixmaster oscillating regions. 

A typical value of the Lyapunov exponents for Mixmaster orbits is (see, e.g., \cite{berger}) $\lambda\sim 0.45$,
so that when $\a>0.225,$ we have synchronization between the two spatial Mixmaster points $A,B$.

\section{Lyapunov function sync}
As is further shown in \cite{c1}, there is in fact a Lyapunov function for the dynamics of sync,  for a  general coupling $g(\Omega)$ satisfying a positivity condition, and  under the further condition that,
\be\label{dot}
\Omega'\cdot\nabla g \leq 0,
\ee
where $\Omega'$ is the orbital derivative of $\Omega$. The Lyapunov function is given by,
\be\label{pot1}
V(\Omega_{+},\Omega_{-})=\frac{1}{2}\left(\Omega_+^2+\Omega_-^2\right)+g(\Omega),
\ee
This implies that the state $\Omega=0$ is asymptotically stable, making synced spatial regions globally asymptotically stable in the past direction.

As an example, consider the simplest, uncoupled, case. The sync equations of the dynamics take the form,
\bq
\Omega_{+}'&=&-(2-q)\Omega_+\label{seqns1}\\
\Omega_{-}'&=&-(2-q)\Omega_-\label{seqns2}.
\eq
Then the function,
\be\label{pot}
V(\Omega_{+},\Omega_{-})=\frac{1}{2}\left(\Omega_+^2+\Omega_-^2\right),
\ee
is a Lyapunov function for the dynamics and satisfies,
\be
V\leq V_0 e^{-4\tau}.
\ee
where $V_0$ is a constant. We conclude that Mixmaster regions in inhomogeneous spacetime synchronize exponentially fast.

\section{Phase sync} 
A different characterization of the possibility of cosmological sync in the assumed primordial chaotic state of the universe may be effected when considering any number of oscillating domains having their own frequencies. This leads to different dynamical evolution of their phases, and the question arises as to whether we can have phase sync between them in this situation.

This problem was shown to have an affirmative answer in two steps. In the first step,  \cite{ba20}, Barrow wrote down the  equation  describing the  phase evolution of a spatial Mixmaster point as a response of the system in terms of the mean field amplitude and phase $r,\psi$ respectively,
\be\label{b2}
\theta'=\frac{3}{\rho}\sqrt{S_+^2+S_-^2}\sin(\theta-\psi),\quad\tan\psi=S_-/S_+,
\ee
with $\rho$ being the polar coordinate in the pair $(\rho,\theta)$ in the $(\Sigma_+,\Sigma_-)$ plane. A further elaboration using these dynamical equations shows that sync is indeed possible.

It can be further shown \cite{c1} that this equation leads to a direct coupling between the oscillating domains, precisely given by a time-dependent Kuramoto-like coupling of spontaneous synchronization. For any partition of $N$ regions, we can show that the  $i$-th domain responds directly to the  $j$-th domain in the partition via the equation \cite{c1},
\be\label{last}
\theta'_i=\frac{3}{N\rho_i}\sqrt{S_+^2+S_-^2}\sum_{j=1}^{N}\sin(\theta_j-\t_i).
\ee

\section{Physical meaning of sync}
We are led to the following physical interpretation \cite{c1} of the process of sync in the present context. The spatial point $B$ will synchronize with the point $A$ lying in $B$'s past after receiving a signal from $A$. This will be so because  their sync function $\Omega\rightarrow 0$, and sync will become progressively apparent when  their corresponding BKL \cite{bkl1,bkl2,bkl3} parameter $u$ values for $A$ and $B$  become equal  with each other in the sense that the numbers of Kasner epochs and eras for the two evolutions of the  spatial points $A,B$  become identical to each other exponentially fast. Thereafter the two points will start oscillating in perfect unison.

This process will lead to the  simplest possible initial state of the universe, characterized by a perfect sync between its different but correlated spatial regions. This will be so because, according to the results reported here, every spatial point will continually send and receive signals from other spacetime points shifting their state of oscillation, adjusting them to that of the other points, and resetting their motion, until sync organizes them all in perfect harmony. In turn, this may have important implications some of which will be discussed elsewhere.

\end{document}